\documentclass{aastex61}

\newcommand\aastex{AAS\TeX}

\usepackage{rotating}
\received{November 27, 2017}
\revised{xxxx xx, 201x}
\accepted{\today}
\submitjournal{ApJ}

\shorttitle{\aastex\ Technique renovation for the GS-reconstruction}
\shortauthors{Li et al.}

\begin{document}

\title{Corner singularity and its application in regular parameters optimization:  technique renovation for Grad-Shafranov reconstruction}

\correspondingauthor{Huijun Li}
\email{hjli@spaceweather.ac.cn}

\author[0000-0003-3493-8112]{Huijun Li}
\altaffiliation{College of Astronautics, Nanjing University of Aeronautics and Astronautics, Nanjing, 210019, China.}
\affil{College of Astronautics, Nanjing University of Aeronautics and Astronautics, Nanjing, 210019, China.}
\affiliation{College of Meteorology and Oceanography, PLA University of Science and Technology, Nanjing, 211101, China.}
\affiliation{SIGMA Weather Group, State Key Laboratory for Space Weather, Center for Space Science and Applied Research, Chinese Academy of Sciences, Beijing, 100018,  China.}

\author{Chongyin Li}
\affiliation{College of Meteorology and Oceanography, PLA University of Science and Technology, Nanjing, 211101, China.}
\affiliation{State Key Laboratory of Numerical Modeling for Atmospheric Science and Geophysical Dynamics, Institute of Atmospheric Physics, Chinese Academy of Sciences, Beijing, China.}

\author{Xueshang Feng}
\affiliation{SIGMA Weather Group, State Key Laboratory for Space Weather, Center for Space Science and Applied Research, Chinese Academy of Sciences, Beijing, 100018,  China.}

\author{Jie Xiang}
\affiliation{College of Meteorology and Oceanography, PLA University of Science and Technology, Nanjing, 211101, China.}

\author{Yingying, Huang}
\affiliation{College of Meteorology and Oceanography, PLA University of Science and Technology, Nanjing, 211101, China.}
\affiliation{School of Electronic Information, Wuhan University, Wuhan, China.}

\author{Shudao Zhou}
\affiliation{College of Meteorology and Oceanography, PLA University of Science and Technology, Nanjing, 211101, China.}



\begin{abstract}

Further studies on the corner singularity of GS reconstruction are are compiled in this paper. It's focused on solution of the Data Completion (DC) problem with the Extended Hilbert Transform (EHT) over plane rectangular region. Optimal selections of the regular parameters in{\it Tikhonov} solution of corresponding DC problem are developed in this study. The 6-parameter regular solutions and the {\it Jacobian} matrix and one {\it Hessian} tensor to the regular parameters are derived in this work. A concise formula for EHT in the near field of corners, which shows property for EHT near the corner, are also provided. It serves as the additional constraints for our parameter optimization problem (OP). Third, a nonlinear convex function defined by the regular solution and the corner constraints is introduced and is taken as the object function for the OP of the 6 regular parameters on half-space $(\mathbf{p}>0)$. Given an initial guess of $\mathbf{p}_0$, the optimal parameters are solved from the OP through a well known constrained nonlinear optimization method. Last, the benchmark tests to the proposed solution approach are carried out, and detailed results from totally 9 different bench-cases are tabulated. In contrast to solutions with given regular parameters, our bench results demonstrate that an objective way for selection of the optimal $\mathbf{p}$ is successfully laid out here. Robustness and efficiency of the suggested new approach are also highlighted in this study. 

\end{abstract}




\section{Introduction} \label{sec:intro}

For observational study on evolution of the flux rope structures in space plasma, a tool for structure reconstruction based on single spacecraft data, especially, a tool with enough accuracy (\cite{Li2013, Li2017}), plays an important role. The tools are usually developed under the stationary assumption ($\partial/\partial t \approx 0$), {\it e.g.}, the magneto-hydrostatic equilibrium (\cite{Grad1958}, \cite{Shaf1958}), the magneto-hydrodynamic equilibrium (\cite{Sonn06}),  {\it etc.}, or even under the slowly evolution assumption ($\partial/\partial t \ll 1$) (\cite{Sonn10}).  In case of static equilibrium, the most widely accepted tool is the GS reconstruction (\cite{Sonn1996}, \cite{Hau1999}, \cite{Hu2001, Hu2003}), however, high accuracy solver for the Grad-Shafranov (GS) equations with insufficient boundary data has seldom been discussed (\cite{Li2017}). Uncertainties exist there, {\it e.g.}, in orientation inversion of the invariant axis for flux ropes, in the fitting of the $Pt(A)$ curve,  in effects of the large impact parameters, or in effects of the ill-posedness of the solution approach, {\it etc.}, in almost every link in technique of the GS reconstruction,  errors are brought into the final numerical results, then erode the recovered structures we want (for details see a recent review of \cite{Hu2017} ).

Ill-posedness of current solution approach within the technique of GS reconstruction has been mentioned by \cite{Gon2015}, and a thorough investigation for this problem has also been presented by \cite{Li2013}, where the solution approach is replaced by a {\it Tikhonov} regularizing scheme with the application of  Hilbert Trasnform (HT) over the plane circle. Theoretical prediction has told us that given enough {\it Cauchy} data, the new solution approach can be well-posed, although how many Cauchy data is sufficient is still an open question (\cite{Li2013, Li2015b}). By using the linearity of the elliptic operator of the GS equations, the solution can be divided into two parts: one is solved from a semi-linear elliptic equation with an homogeneous {\it Dirichlet} boundary condition, and the other one is solved from the DC problem to the {\it Laplace's} equation, which is the essential idea of our series study on technique renovation for GS reconstruction (\cite{Li2014a, Li2014b, Li2016a}). Since the solution approaches for the homogenous semi-linear part are discussed maturely in literatures, the only nontrivial task in solution of GS equation is thus reduced to the solution of DC problem of the {\it Laplace's} equation. 

In contrast to traditional treatments for ill-posedness of the so-called elliptic {\it Cauchy} problems (\cite{Gupta2009, Gupta2012}), we treat the main obstacle of the ill-posedness by reducing them into the DC problem of Laplace's equation, where the missing data are completed with the HT formulae \citep{Li2013}. Since the considered problem is essentially of elliptic type (well-posed one), and thus we need only to recover the missing boundary data. This becomes a task to solve DC problem on the Sobolev space \citep{Yu2006, Li2013}. Combining with application of HT formulae not only take a chance to conquer the ill posedness, but also produce a solution with very high accuracy.  The high accuracy property for HT over the circular boundary has also been reproduced after removal of the Cauchy singularities \citep{Li2015b}, which validates the theory prediction that `given enough Cauchy data, the DC problem is well-posed'.  

We have also carried out this idea in mountain gravity wave reconstruction \citep{Li2015a} and its near-field computation \citep{Li2016b}.  An idea about extending the HT to that over the plane rectangular region has also been reported in an abstract by \cite{Li2014b}.  Full approach for GS reconstruction has been reported in an abstract by \cite{Li2014a}, and then presented in a talk by \cite{Li2016b}. All studies along the line have shown a broad application prospective of this idea. In order to carry out it in a more realistic context, the HT over plane circle is extended to that over the plane rectangle \citep{Li2017}. The ill posedness for the essential technique of GS reconstruction are solved in \cite{Li2017}. New data completion approach is built with these EHT formulae, and a new three-parameter regularization scheme is developed to get a stable solution from the first-kind {\it Fredholm} system. Numerical experiments are also carried out with the analytic solutions. Bench tests results from the forward computation and its reversion highlights its efficiency and accuracy (\cite{Li2017}).  However, it is only the first step in the long route to the technique renovation, although the new approach takes us a chance to solve the major obstacle, {\it i.e.}, the ill-posedness for the GS technique.  

As discussed by \cite{Li2017}, there are still five major questions about the new approach for DC problem over rectangular domain, which are listed here again for completeness:
\begin{description}
\item [1] Can the DE rule for integral with both end-point singularities be replaced by the trapezoidal rule for more efficiency? 
\item [2] Can we get rid of the dependence on SC tool by \cite{Torbin2002}, and replace it with any other analytic solutions ({\it i.e.}, elliptic functions)?  
\item [3] How to define the Hilbert transform at the four corners, and how to give a definite computation for its HT results? 
\item [4] How to select the regularization parameters (like, $\alpha$, $\beta$, and $\gamma$) in an objective way? 
\item [5] How to get the new quadrature scheme with needed accuracy, when oscillations exist in the boundary gradients ?
\end{description}

We focus on Question (3), and (4) in this study, {\it i.e.}, the EHT near the corner, and the optimal selection of the regular parameters. The regular parameters are currently given in a objective way, which is not suitable for running it automatically in our final approach. To our knowledge, how to get an optimal selection for the regular parameters is also an active topic in the community of inversion problem. More than this, EHT near the corner of the rectangular region has not been discussed yet. In this paper, the regularized 6-parameter solution is derived explicitly in Section \ref{OPTG}, and the EHT formulae in the near field of corners are derived in Section \ref{EHTC}. Then, the nonlinear object function defined by the regular solutions and the corner constraints are presented in Section \ref{OP}, and we minimizing the object function through the selected nonlinear constraint optimization method. Lastly, we tabulate benchmark testing results in Section \ref{BCT}, where the optimal parameters and corresponding errors for 9 different cases are presented in contrast to corresponding analytic solutions. Discussions and conclusions are presented in Section \ref{SUMMC}.

\section{6-parameter regular solution}
\label{OPTG}
The linear system in Equation (23) \citep{Li2017} can be formulated, for each component of the boundary gradients, as follows: 
\begin{eqnarray}
\mathbf{A}\mathbf{x}_{\rm n} + \mathbf{B}\mathbf{y}_{\rm n} + \mathbf{C}\mathbf{z}_{\rm n} &=& \mathbf{f}_{\rm n} \nonumber\\
\mathbf{A}\mathbf{x}_{\rm t} + \mathbf{B}\mathbf{y}_{\rm t} + \mathbf{C}\mathbf{z}_{\rm t} &=& \mathbf{f}_{\rm t}\nonumber
\end{eqnarray}
where, the subscripts ${\rm n}, {\rm t}$ denotes the normal and tangential components, respectively. $\mathbf{f}_{\rm n}$, and $\mathbf{f}_{\rm t}$ are column vectors, which denote items at right side of the {\it Fredholm} system in Equation (23) \citep{Li2017}. By introducing three regular parameters for each component of the regular solution, we can get the final regular solution with totally 6 unknown parameters. Let ($\mathbf{v}_{\rm n}, \mathbf{v}_{\rm t}$) denotes the unknown boundary gradients components, and then we formulate the linear system as follows:
\begin{equation}
\left[\begin{array}{cccccc}
\mathbf{A} & \mathbf{B} & \mathbf{C} &  &   &   \\
  &   &   & \mathbf{A} & \mathbf{B} & \mathbf{C}  
\end{array} \right] \left[\begin{array}c \mathbf{v}_{\rm n} \\ \mathbf{v}_{\rm t} \end{array}\right]
= \left[\begin{array}c \mathbf{f}_{\rm n} \\ \mathbf{f}_{\rm t} \end{array}\right]\nonumber
\end{equation}
 where, $\mathbf{v}_{\rm n} = [\mathbf{x}_{\rm n}; \mathbf{y}_{\rm n}; \mathbf{z}_{\rm n}]$, and $\mathbf{v}_{\rm t} = [\mathbf{x}_{\rm t}; \mathbf{y}_{\rm t}; \mathbf{z}_{\rm t}]$, both are column vectors. 
 
Let $\mathbf{M}$ denote the coefficient matrix in blocks of  $\mathbf{A}$, $\mathbf{B}$, and $\mathbf{C}$,  and $\mathbf{v} = [\mathbf{v}_{\rm n}; \mathbf{v}_{\rm t}]$, $\mathbf{b} = [\mathbf{f}_{\rm n}; \mathbf{f}_{\rm t}]$. Let $\mathbf{p}$ denote the introduced 6 regular parameters, a column vector of $\mathbf{p} = [p_1; p_2; p_3; p_4; p_5; p_6]$. Then we can define a convex function as follows:
\begin{equation}
\begin{array}{ccl}
f(\mathbf{v}; \mathbf{p}) &=& ||\mathbf{M} \mathbf{v} - \mathbf{b}||^2 \\&+& p_1 ||\mathbf{L}\mathbf{x}_{\rm n}||^2 + p_2 ||\mathbf{L}\mathbf{y}_{\rm n}||^2 + p_3 ||\mathbf{L}\mathbf{z}_{\rm n}||^2 \\&+& p_4 ||\mathbf{L}\mathbf{x}_{\rm t}||^2 + p_5 ||\mathbf{L}\mathbf{y}_{\rm t}||^2 + p_6 ||\mathbf{L}\mathbf{z}_{\rm t}||^2.
\end{array}
\end{equation}
where, $p_i>0, i=1,\cdots, 6$. The {\it Tikhonov} solution $\mathbf{v}$ can be solved from minimization of this convex function, {\it i.e.}, $\mathbf{v} = \arg\min\{f(\mathbf{v}; \mathbf{p})\}$, which can be written as follows:
\begin{equation}
\mathbf{v}(\mathbf{p}) = \mathbf{R}^{-1} \mathbf{M}^T \mathbf{b}, \label{TS}
\end{equation}
where, $\mathbf{R} = \mathbf{M}^T\mathbf{M} + {\rm diag}( p_1\mathbf{L}^T\mathbf{L}, \cdots,  p_6\mathbf{L}^T\mathbf{L} )$ is a linear function of regular parameters $\mathbf{p}$. The operator $\rm diag (\cdot)$ builds a dialog matrix with blocks within the brackets. This is the 6-parameter regular solution, a nonlinear function of the regular parameter of $\mathbf{p}$, which can be taken as an extension of the 3-parameter one in \citep{Li2017}.

Let $(\mathbf{L}^T \mathbf{L})_{i}$, ($ i=1,\cdots, 6$) denote the block matrix with all blocks are zeros, but for the $i^{th}$ block along the dialog, which equals to $\mathbf{L}^T \mathbf{L}$. Then the {\it Jacobian} of $\mathbf{v}(\mathbf{p})$ can be written as follows:
\begin{equation}   
\mathbf{J}(\mathbf{v}; \mathbf{p}) = -\left[\mathbf{R}^{-1}\left(\mathbf{L}^T\mathbf{L}\right)_1\mathbf{v}, \cdots, \mathbf{R}^{-1}\left(\mathbf{L}^T\mathbf{L}\right)_6\mathbf{v} \right]. \label{JV}
\end{equation}
It is a matrix composited by 6 column of vectors. 

The {\it Hessian} tensor for $\mathbf{v}(\mathbf{p})$ can be written as follows:
\begin{equation}  
\mathbf{H}(\mathbf{v}; \mathbf{p}) =  \left[\begin{array}{ccc} 
\mathbf{R}^{-1}\mathbf{H}_{11} \mathbf{v} & \cdots & \mathbf{R}^{-1}\mathbf{H}_{16} \mathbf{v}\\  
\vdots & \ddots & \vdots \\
\mathbf{R}^{-1}\mathbf{H}_{61}\mathbf{v} & \cdots &  \mathbf{R}^{-1}\mathbf{H}_{66}  \mathbf{v}
\end{array} \right] , \label{HV}
\end{equation}
where, $\mathbf{H}_{ij} = \left(\mathbf{L}^T\mathbf{L}\right)_i \mathbf{R}^{-1} \left(\mathbf{L}^T\mathbf{L}\right)_j$, and $i,j = 1, \cdots, 6$. This {\it Hessian} is a 3-order tensor composited by $6\times 6$ column vectors.  

Lastly, with help of the {\it Jacobians} and {\it Hessians}, the nonlinear regular solution of $\mathbf{v}(\mathbf{p})$ can now be expanded within the field near $\mathbf{p}$ as follows:
\begin{equation}   
\mathbf{v}(\mathbf{p} + \delta \mathbf{p}) = \mathbf{v}(\mathbf{p}) + \mathbf{J}(\mathbf{v}; \mathbf{p}) \delta\mathbf{p} + \frac{1}{2}\delta \mathbf{p}^T \mathbf{H}(\mathbf{v}; \mathbf{p}) \delta \mathbf{p} + {\rm o} (||\delta \mathbf{p}||^3). 
\end{equation}
After truncation, the result can serve as a second order approximation to the regular solution with $|\delta \mathbf{p}|<1$.

\section{EHTs near corner of the rectangle}
\label{EHTC}

As shown by Equation (7), and (8) in \cite{Li2017}, formulae of Hilbert transform over the circular boundary are extended into a couple of boundary integrals over the plane rectangle, where the circular boundary of $\Gamma$ is conformal mapped from the plane rectangle. However, the path of these two boundary integrals (EHTs) are defined by $\Gamma' = \Gamma\setminus\{x_k\}$, ($k=1,2,3,4$). It indicates that there are four singular points mapped from corners of rectangle onto the integral path $\Gamma$, and they are discarded from $\Gamma$ in the final integral formulae. Although EHT does not exist at the singular corner positions in context of the plane rectangle, the EHT relations between gradient components do exist near the four singular positions. They are important properties for EHT near the singular corner of plane rectangle, which are ignored in our previous study. As shown by the following sections in this paper, it plays an important role in optimal selection of the regular parameters for solution to the corresponding DC problem. 

\subsection{Properties for EHT in near field of corners}
We present a detailed derivation of  the explicit formulae for relations of the EHTs near the four singular corners, {\it i.e.}, $x_k$, ($k=1, 2, 3, 4$). As the conformal mapped boundary gradient components of $(g_n, g_t)$ can be expressed with the Cartesian components $(u_x, u_y)$ in an explicit way \citep{Li2017}, we can rewrite it in the following matrix form:
\begin{equation}
\left[
\begin{array}c
g_n(s) \\ g_t(s)
\end{array}
\right] = \left[
\begin{array}{cc}
C_{\rm r}(s) & -C_{\rm i}(s) \\
C_{\rm i}(s) & C_{\rm r}(s)
\end{array}
 \right]\left[ 
\begin{array}c
u_x(s) \\
u_y(s)
\end{array}
\right], \label{mgtgn} 
\end{equation}
where $C_{\rm r} (s), C_{\rm i}(s)$ are the real and imaginary part, respectively, which can be written as follows:
$$C_{\rm r}(s) + i C_{\rm i}(s) =  - \frac{C \exp(i\lambda)}{I(s)}, $$
where, $s\in \Gamma'$, $C$ is the mapping constant, $i=\sqrt{-1}$,  $\lambda = \frac{1}{4}\sum_{k=1}^4 x_k$, and the sign item $I(s)$ can be written as follows:
$$I(s) = \prod_{k=1}^4 \sqrt{{\rm sign}(s-x_k)}.$$
At those corners, say $s=s_k$ $(k=1,2,3$, and $4)$,  $I(s) = 0$, then $g_n(s)$, $g_t(s)$ become singular, and thus no definition for $g_n, g_t$ at $s_k$ are given in our previous study \citep{Li2017}. But within the near field of $s_k$, where a concise form of $g_n, g_t$  do exist. Unfortunately, the EHT relations for them have not been discussed yet in \cite{Li2017}. 

Because the Cartesian components $(u_x, u_y)$ in physical space has only one value at corner, so that there exists the relation between the values of $(u_x, u_y)$ at the left of the corner and the one at the right, as both of them approach to the corner position:
\begin{equation}
\lim_{s \to s_k^{\pm}}\left[ 
\begin{array}c
u_x\\
u_y
\end{array}
\right](s) = \left[ 
\begin{array}c
u_x\\
u_y
\end{array}
\right](s_k). \label{EHTC1}
\end{equation}

As  the sign function ${\rm sign}(s)$ has limitations, {\it i.e.}, $\lim_{s\to x_k^{-}} {\rm sign}(s-x_k) = -1$, and $\lim_{s\to x_k^{+}} {\rm sign}(s-x_k) = 1$, then there exists the relation between sign items near the corner, {\it i.e.}, $I(s_k^+) = i I(s_k^-)$, and thus it can be proved that there exists the relation between the components of $(g_n, g_t)$ near the corner:
\begin{equation}
\left[ 
\begin{array}c
g_n\\
g_t
\end{array}
\right](s_k^+) = \mathbf{H} \left[
\begin{array}c
g_n\\
g_t
\end{array}
\right](s_k^-),
\end{equation}
where, $\mathbf{H} =  [ 0 , -1; 1 , 0 ]$, a sympletic unitary matrix,  which establishes the EHT relation for $(g_n, g_t)$ near the corner. 

These constraints for $(g_n, g_t)$ near corner has never been discussed in previous study, which can be used as the main constraints in optimal selection of the regular parameters in this study.

\subsection{Corner constraints}
Corner constraints in this study can be divided into two categories, {\it i.e.}, one is the direct constraints, which are produced by transformation with Equation (\ref{EHTC1}) to the known {\it Cauchy} data at both end of the {\it Cauchy} line, {\it i.e.}, $[\mathbf{g}_{\it t}; \mathbf{g}_{\rm n}](1, \rm{end})$. The other is the indirect constraints, which are produced by EHT to the optimal solutions at corresponding end of the opposite side, {\it i.e.}, $[\mathbf{\hat{x}}_{\rm t}; \mathbf{\hat{x}}_{\rm n}]({\rm end})$, and $[\mathbf{\hat{z}}_{\rm t}; \mathbf{\hat{z}}_{\rm n}](1)$. Both kind of corner constraints can be used as the constraint conditions for parameter optimizations. 

For solution $\mathbf{\hat{x}}$, one additional constraint can be produced with EHT within the near field of the corner between side $\mathbf{\hat{x}}$ and the {\it Cauchy} line. This constraint can be formulated as follows:
\begin{equation}
\left[\begin{array}c \mathbf{\hat x}_{\rm t}\\\mathbf{\hat x}_{\rm n}\end{array}\right](1) = \mathbf{H}\left[\begin{array}c \mathbf{g}_{\rm t}\\\mathbf{g}_{\rm n} \end{array} \right]({\rm end}), \label{CX}
\end{equation}
 where, `1' and `${\rm end}$' in the parenthesis, denote the indices for the first, and the last element at corresponding column vector, respectively. For solution $\mathbf{\hat{z}}$, another constraint can be produced with following EHT transformations within the near field of the corner between $\mathbf{\hat{z}}$ and  the {\it Cauchy} line, which can be formulated as follows:
\begin{equation}
\left[\begin{array}c \mathbf{\hat z}_{\rm t}\\\mathbf{\hat z}_{\rm n}\end{array}\right]({\rm end}) = \mathbf{H}^{-1}\left[\begin{array}c \mathbf{g}_{\rm t}\\\mathbf{g}_{\rm n} \end{array} \right](1) , \label{CZ}
\end{equation}
where $\mathbf{H}^{-1} = -\mathbf{H}$. 

As for the indirect constraints at both ends of the opposite side, solution $\mathbf{\hat{y}}$ can be constrained at both ends, and corresponding constraints are formulated as follows:
\begin{equation}
\left[\begin{array}c \mathbf{y}_{\rm t}\\\mathbf{y}_{\rm n}\end{array}\right](1) = \mathbf{H}\left[\begin{array}c \mathbf{\hat{x}}_{\rm t}\\\mathbf{\hat{x}}_{\rm n} \end{array} \right]({\rm end}),  \label{CY1} \\
\end{equation}
and 
\begin{equation}
\left[\begin{array}c \mathbf{y}_{\rm t}\\\mathbf{y}_{\rm n}\end{array}\right]({\rm end}) = \mathbf{H}^{-1}\left[\begin{array}c \mathbf{\hat{z}}_{\rm t}\\\mathbf{\hat{z}}_{\rm n} \end{array} \right](1). \label{CY2}
\end{equation}

Lastly, we get the totally 4 different linear systems (a total of 8 equations), however, there are only 6 unknown parameters from which to be solved, which indicates that it is an over determined system, and the unknown regular parameters should be solved in a sense of least square.

By introducing the end point operator $\mathbf{s} = [1, 0, \cdots, 0]$, and $\mathbf{e} = [0, \cdots, 0, 1]$, and thus there exists the relations, {\it e.g.}, $\mathbf{s} \mathbf{x}_{\rm n} =  \mathbf{x}_{\rm n}(1)$, $\mathbf{e} \mathbf{x}_{\rm n} =  \mathbf{x}_{\rm n}({\rm end})$. The corner constraints can be reduced into a matrix form:
\begin{equation}
\mathbf{X} \mathbf{v}(\mathbf{p}) = \mathbf{c} \label{CC}
\end{equation}
where, the coefficient matrix are composed by the end point operators:
$$
\mathbf{X} = \left[\begin{array}{cccccc} 
\mathbf{s}  & \mathbf{0} & \mathbf{0} & \mathbf{0} & \mathbf{0} & \mathbf{0}\\
\mathbf{0} & \mathbf{s}  & \mathbf{0} & \mathbf{e} & \mathbf{0} & \mathbf{0}\\
\mathbf{0} & \mathbf{e} & \mathbf{0} & \mathbf{0} & \mathbf{0} & -\mathbf{s}\\
\mathbf{0} &\mathbf{0} & \mathbf{e} & \mathbf{0} & \mathbf{0} & \mathbf{0} \\
\mathbf{0} & \mathbf{0} & \mathbf{0} & \mathbf{s} & \mathbf{0} & \mathbf{0}\\
-\mathbf{e} & \mathbf{0} & \mathbf{0} & \mathbf{0} & \mathbf{s} & \mathbf{0}\\
\mathbf{0} & \mathbf{0} & \mathbf{s} & \mathbf{0} & \mathbf{e} & \mathbf{0}\\
\mathbf{0} &\mathbf{0} &\mathbf{0} & \mathbf{0} &\mathbf{0} & \mathbf{e} 
\end{array}\right], 
$$
where, $\mathbf{0}$ is a row of zeros. The column vector $\mathbf{c}$ can be formulated as follows:
$$
 \left[-\mathbf{e} \mathbf{g}_{\rm t}; 0; 0; \mathbf{s} \mathbf{g}_{\rm t}; \mathbf{e} \mathbf{g}_{\rm n}; 0; 0; -\mathbf{s} \mathbf{g}_{\rm n} \right],
$$
where, $\mathbf{g}_{\rm n}$, and $\mathbf{g}_{\rm t}$ are the known {\it Cauchy} data along the {\it Cauchy} side.

\section{Parameter optimization}
\label{OP}
In the recovered boundary gradient components, as plotted by Figure (3) in \cite{Li2017}, we find that the nearer the position to both end of the {\it Cauchy} line and to that of the opposite side, the bigger the errors there be. These facts inspired us to control those errors with the corner constraints discussed in previous section.  A convex function is built with the corner constraints in Equation (\ref{CC}), which is formulated as follows:
\begin{equation}
f(\mathbf{v}; \mathbf{p}) = \left[\mathbf{X} \mathbf{v}(\mathbf{p}) - \mathbf{c}\right]^T\left[\mathbf{X} \mathbf{v}(\mathbf{p}) - \mathbf{c}\right], \label{OBF}
\end{equation}
where, $T$ denotes the transpose. Obviously, the optimal parameter $\mathbf{p}$ that satisfied the corner constraints in Equation (\ref{CC}), must also produce a minimal value of this convex function, which is the so-called object function. That's a nonlinear function, and the optimal parameters $\mathbf{p}$ can be solved through the nonlinear minimization of $f(\mathbf{v}; \mathbf{p})$. The procedure can be expressed by the following formula:
\begin{equation}
\arg\min_{\mathbf{p}} f(\mathbf{v}; \mathbf{p}), \label{NOP}
\end{equation}
where, parameter $\mathbf{p}$ should be searched in half space of $\mathbf{p}>0$. A lot of nonlinear minimization algorithm exists in literature, which are suitable for solving the OP in Equation (\ref{NOP}). We select here the so-called 'trust-region-reflective' algorithm for nonlinear OP (\cite{ST1983}). According to this algorithm, we take a small modification to OP (\ref{NOP}), with the linear inequal constraints of $\mathbf{p}>0$ replacing by the bound constraints of $\epsilon \mathbf{I}<\mathbf{p}< \mathbf{I}/\epsilon$, where the column vector $\mathbf{I} = [1; \cdots; 1]$. 

In order to improve the accuracy and efficiency of the minimization procedure, explicit formulae of the {\it Jacobians} and {\it Hessians} of $f(\mathbf{v}; \mathbf{p})$ are prepared for the selected algorithm  (\cite{ST1983}). The {\it Jacobians} for  $f(\mathbf{v}; \mathbf{p})$ can be written as follows:  
$$
\mathbf{J}(f; \mathbf{p}) = 2 \mathbf{J}^T(\mathbf{v}; \mathbf{p}) \mathbf{X}^T(\mathbf{X} \mathbf{v} - \mathbf{c}),
$$
where, $T$ denotes transpose, and $\mathbf{J}(\mathbf{v}; \mathbf{p})$ is defined by Equation (\ref{JV}).
The {\it Hessians} for  $f(\mathbf{v}; \mathbf{p})$ can be written as follows:    
$$
\mathbf{H}(f; \mathbf{p}) = 2 \mathbf{J}^T(\mathbf{v}; \mathbf{p}) \mathbf{X}^T\mathbf{X} \mathbf{J}(\mathbf{v}; \mathbf{p}) + 2 \mathbf{H}^T(\mathbf{v}; \mathbf{p}) \mathbf{X}^T(\mathbf{X}\mathbf{v} - \mathbf{c}),
$$
where, $\mathbf{H}(\mathbf{v}; \mathbf{p})$ is defined by Equation (\ref{HV}).

\section{The bench-case tests}
\label{BCT}
Bench-case tests for the proposed approach are carried out in this study. As listed in Table (\ref{AS}), totally 9 analytic harmonic solutions are selected from literatures, {\it e.g.},  \cite{Gupta2009}, and \cite{Gupta2012}. Their Cartesian gradients are also derived explicitly and are listed in Table (\ref{AS}), which are used to produce the {\it Cauchy} data and the exact solutions for the recovered boundary gradients.  These 2D functions in each case are harmonic on full plane ($R^2$), however, in order to avoid symmetries of them in the domain selected in previous paper \citep{Li2017}, we move the domain to $[1-i, 3-i, 3+i, 1+i]$ on the complex plane, and let {\it Cauchy} line defined by $[1-i, 3-i]$. The mapping constants, $C$, is the same as that in our previous paper, for details about $C$ please refer to \cite{Li2017}. To save the computation time in each iteration,  we set the parameter $M=30$ for meshing of the {\it Cauchy} line, $N = 45$ for meshing in the DE rules, and $P=5$ for approximation with the  {\it Chebyshev} series. The three parameters are selected smaller enough to show the efficiency of the proposed optimization method.

\begin{table}
\caption{The selected plane harmonic functions $u(x,y)$ and its Cartesian gradients [$u_x, u_y$]}
\begin{center}
\begin{tabular}{c|c|c|c} \label{AS}
 Case (\#) &$u(x, y)$ & $u_x$ & $u_y$ \\
\hline
1&$-y$ & 0 & -1\\
2&$x$ & 1 & 0 \\  
3&$xy$ & $y$ & $x$\\
4&$x^2-y^2$ & $2x$ &$-2y$\\  
5&$x^3-3xy^2$ & $3x^2 - 3y^2$ & $-6xy$\\
6&$y^3-3x^2y$ & $-6xy$ & $3y^2-3x^2$\\  
7&$x^3-6x^2y-3xy^2+2y^3$ & $3x^2 - 12xy - 3y^2$ & $-6x^2 - 6xy + 6y^2$\\
8&$\sin x  \sinh y$& $\cos x  \sinh y$& $\sin x \cosh y$\\
9&$\cos x \cosh y$ & $-\sin x \cosh y$ & $\cos x \sinh y$\\
\hline 
\end{tabular}
\end{center}
\label{default}
\end{table}%

\begin{figure*}     
\gridline{\fig{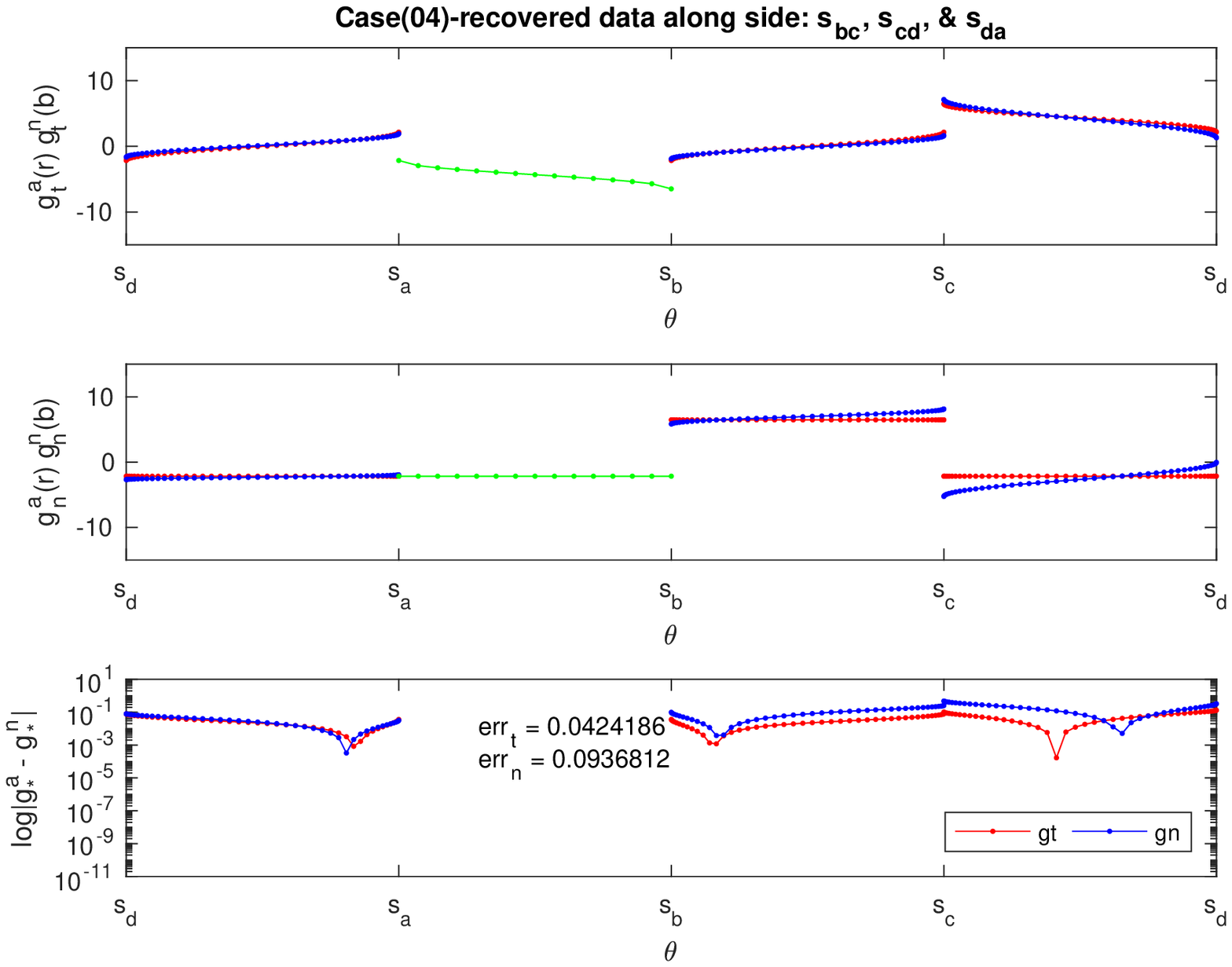}{0.3\textwidth}{(a)}
              \fig{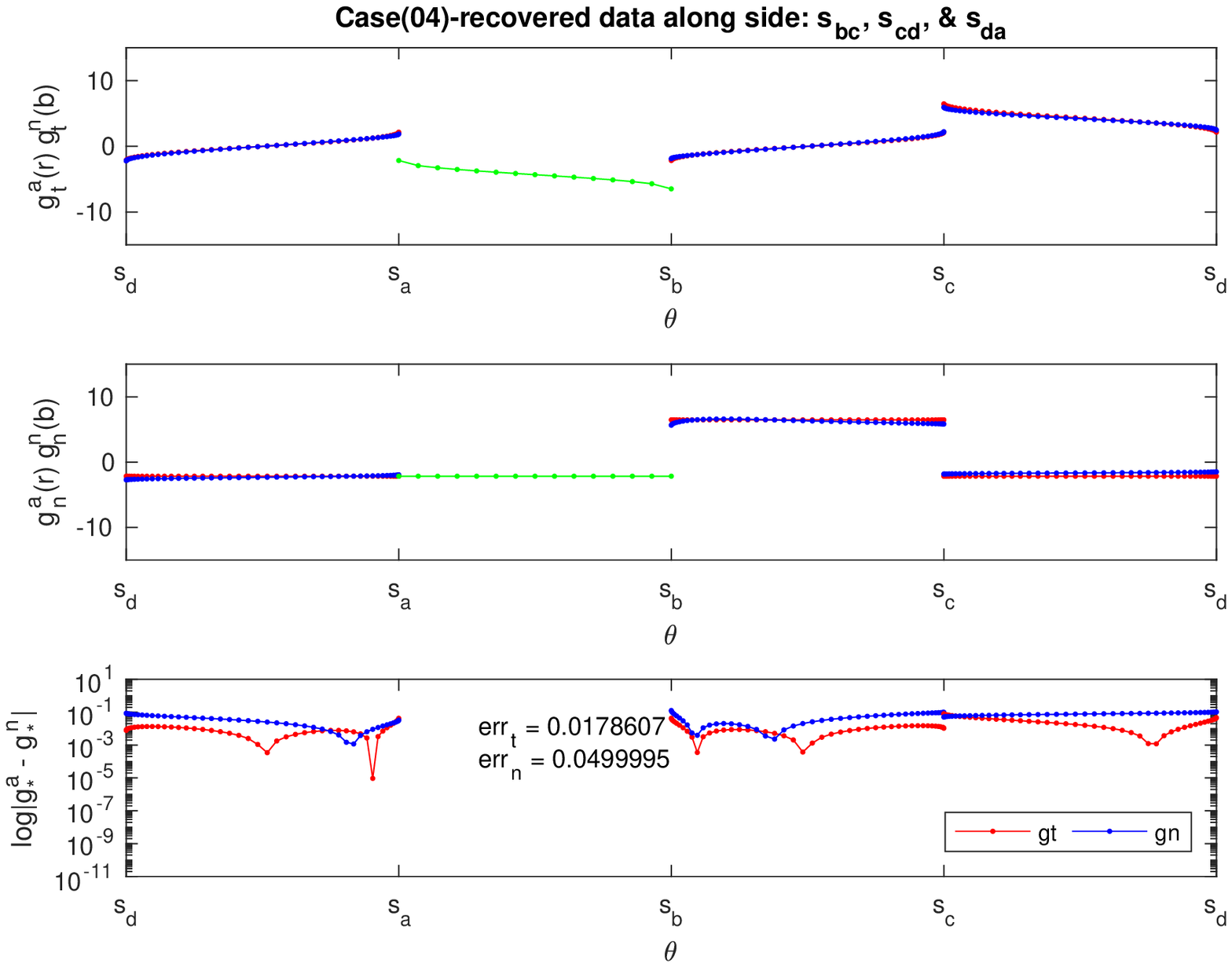}{0.3\textwidth}{(b)}
              \fig{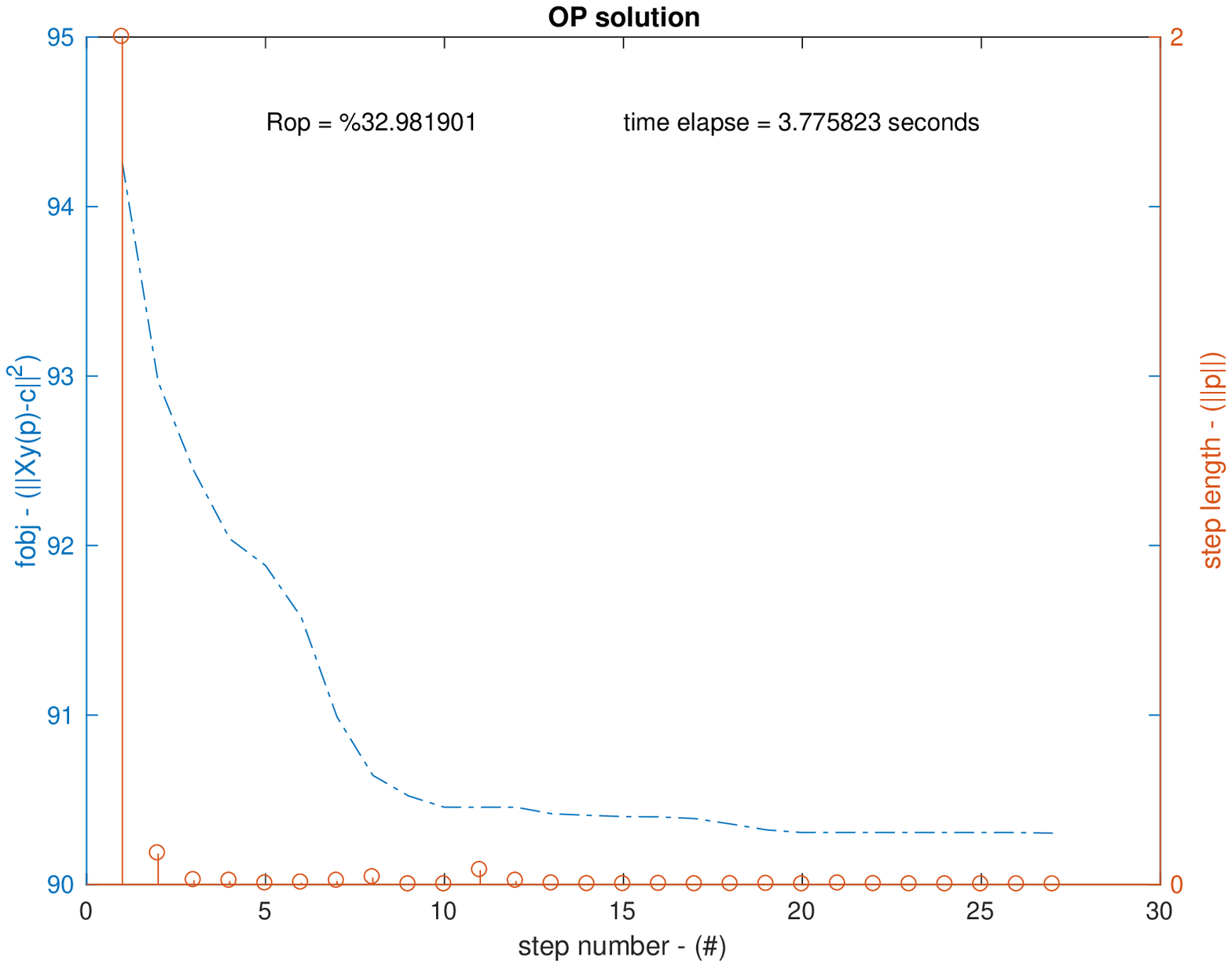}{0.4\textwidth}{(c)}}
\caption{Plots of the recovered data given $\mathbf{p}_0 = [1;1;1;1;1;1]$ (left), and with the optimal parameter of $\mathbf{p} = [0.000002; 1.535130; 0.007792; 0.000000; 0.985085; 0.000001]$ (middle). Detail descriptions of these plots, see \citep{Li2017}. The optimization procedure are also shown by plots on the right, where vale of the object function $f_{\rm obj}$ vs. the iteration steps (the decreasing dash-dot line), and the step length of $\delta \mathbf{p}$ vs. iteration steps (the stem plot) are plotted. The optimal ratio $\rm{R_{\rm op}} = |\Delta f_{\rm obj}|/f_{\rm obj}$ and the total time elapse are also shown by texts.
\label{CASE}}
\end{figure*}

A sample plots for numerical results from Bench-case ($4$) are shown by Figure (\ref{CASE}). The recovered boundary gradient components with the given regular parameters of $\mathbf{p}_0 = [1;1;1;1;1;1]$ are computed directly from the 6-parameter {\it Tikhonov} solution in Equation  (\ref{TS}). As shown by plots on the left, the mean errors in recovered data are ($\%4.24186, \%9.36812$) for the tangent and normal components, respectively. After optimization, the optimal regular parameter becomes $\mathbf{p} = [0.000002; 1.535130; 0.007792; 0.000000; 0.985085; 0.000001]$, and the new results are plotted on the middle, where the mean errors for the recovered data become ($\%1.78607, \%4.99995$). It is decreased obviously in contrast to the pre-optimization couple. As plotted on the right of Figure  (\ref{CASE}), totally 27 iterations are there for the optimization procedure, and values of the object function $f(\mathbf{v}; \mathbf{p})$ are decreased step by step, until the normal of step became smaller than the tolerance of $1\times10^{-6}$. The optimal ratio is $R_{\rm op} = \%32.981901$, and the computation cost is 3.775823 seconds. It indicates that the proposed optimization approach have an efficient effects on objective selection of  the regular parameter $\mathbf{p}$. 

The same results can be found in all the other 8 selected bench-cases. To avoid repeating of the plots, we list the main data of the 9 numeric experiments in Table (\ref{TR}), where $[p_1, p_2, p_3, p_4, p_5, p_6]$ are components of the optimal regular parameter $\mathbf{p}$,  $\overline{err}_{{\rm *}0}$, $\overline{err}_{{\rm *}}$ (* denotes ${\rm t}$ or ${\rm n}$) are the mean errors (in percentage) of the recovered boundary gradients, with subscript of 0 for pre-optimization and without for after-optimization,  $N$ is the total iterations, and $T$ is the time elapse for the total iterations. Because we set the initial value of $\mathbf{p}_0$ in each bench-case test with the same value, saying $[1;1;1;1;1;1]$, although the final optimal regular parameter for each case are different, and the other data are also different for each case, we can still find that the totally mean errors decreased from  ($\%10.2906, \%9.52073$)  to ($\%5.13994, \%4.93273$), and the ratio of optimization reaches  $\%42.60624$ with the mean 17 iterations and an average time cost smaller than 3 seconds. It also indicates that our optimization with the corner constraints of {\it Hilbert} transfroms play a crucial role in optimal selection of the regular parameter $\mathbf{p}$. 

One thing must be explained separately is that as shown in Table (\ref{TR}), there are three values of the optimal parameters, {\it i.e.}, $p_1$ in Case (3), $p_4$ in Case (4) and (9), have the value of $0.000000$, which seems conflict with our inequal constraints of $\mathbf{p}>0$. In facts, there are no conflicts,  because the optimal parameter are searched between [$10^{-9}, 10^4$] in current numeric experiments, as these parameters smaller than $10^{-7}$, they are shown in our outputs by $0.000000$, but still satisfies the request of $\mathbf{p}>0$.

\section{Summary and Conclusion}
\label{SUMMC}
New tools are developed in this study after the corner singularity of GS reconstruction techniques. One is the 6-parameter regular solution for the DC problem over the plane rectangular region, which are introduced for the first time in this work. Another one is the nonlinear constraints build after the new {\it Hilbert} transform relations within the near field of corners. We formulate the DC problem into a nonlinear constraint optimization problem (NCOP) with the help of these two important tools. The optimal regular parameters are solved from this NCOP with the well known NCOP solver. Numerical results are carried out in this study, which show the efficiency of our proposed iterative approach. A objective way for selection of the regular parameters that appeared in solution of the DC problems is declared for the first time in this study.

The 6-parameter regular solution has greatly improved the computation efficiency. In contrast to the 3-parameter regular solutions in previous study, where the solution for tangent and normal components are computed separately, the computation cost greatly decreased after combing them into one matrix operation,  {\it e.g.}, as reported in Table (\ref{TR}), for the average 17 iterations, the cost is no more than 3 seconds.
As shown by Table (\ref{TR}), mean errors for the recovered data are around $\%5$,  {\it i.e.}, an average for the totally 9 bench-cases is  ($\%5.13994, \%4.93273$). It indicates that the proposed method has definitely effects in err control for solution of the DC problems. 

However, as the uniqueness and existence for the DC problem are proofed for the plane harmonic functions, there should be a way to get the exact solution for each DC problem, which means that the mean errors should be controlled to the machine precision. That will take great chance to study on stability transitions of those stationary structures. Where dose the new way lies? That's really a great challenge for our pursuit for the new solver. Although we're still a long way from achieving the final approach, we really build a practicable solver for DC problems, with which we can solve the problem in an objective way for the first time.   

As the 5 major concerns discussed in our previous study, we settled the $3^{\rm rd}$ and the $4^{\rm th}$ question in this study, the $1^{\rm st}$, $2^{\rm nd}$, and $5^{\rm th}$ questions are still open, which need to be addressed in our future study.  

\acknowledgments

The work is jointly supported by the National Natural Science Foundation of China (40904048, 41275029, 41301370, 41375045), the General Financial grant from the China Postdoctoral Foundation (2011M500151), the Special Financial Grant from the China Postdoctoral Science Foundation (2014T70965),  and the National Basic Research Program of China (2012CB825606). The work of J. Xiang was supported by the National Natural Science Foundation of China (41275113). We acknowledge the help from Dr. Xiang, C.-Q., J.-S. Yao, X.-H. Zhao, P.-B. Zuo , J.-P. Guo and Z. W. Jiang, who are fellows in the SIGMA group at CSSAR, and the helpful discussions with Prof. Huang, S. X. from the PLAUST. We are grateful to the anonymous referees for their careful reading of the manuscript that lead to an improved presentation. No data was used in producing this manuscript.

\newpage
\begin{sidewaystable}[h]
\caption{Numerical results for the totally 9 bench-case testings to the proposed iterative approach for optimal regular parameters.}
\tablewidth{400pt}
\centering
\begin{tabular}{c|c|c|c|c|c|c|c|c|c|c|c|c|c}
\hline
Case ($\#$) & $p_1$ & $p_2$ &$p_3$ & $p_4$ & $p_5$ & $p_6$ & $\overline{err}_{{\rm t}0} (\%) $ & $\overline{err}_{{\rm n}0} $ (\%) & $\overline{err}_{\rm t} $ (\%)& $\overline{err}_{\rm n} $ (\%)&  $R_{\rm op}$ (\%) & N & T (s)  \\
\hline
$1$ &2.776118 & 0.906650 & 1.901519 & 0.000023 & 2.222776 & 0.000035 &  17.42789 & 0.80727 & 3.75155 &  0.80727 & 95.180502 & 21 & 3.374148 \\
$2$ &0.000020 & 2.166655 & 0.000004 & 1.651869 & 0.902847 & 1.638288 & 0.80726 & 17.42788 & 0.80726 & 6.09649  &  94.756648 & 20 & 3.143033 \\
$3$ &0.000000 & 0.877659 & 0.000003 & 0.000002 & 1.501945 & 0.011270 & 11.84909 & 4.21700 &  5.52463 & 0.59318 & 33.527140 & 29 & 4.074154 \\
$4$ &0.000002 & 1.535130 & 0.007792 & 0.000000 & 0.985085 & 0.000001 & 4.24186 & 9.36812 & 1.78607 & 4.99995 & 32.981901 & 27 &  3.827881\\
$5$ &0.000002 & 1.107745 & 27.327369 & 41.921196 & 3.835267 &  0.000018 & 6.72510 & 16.22428 & 6.88687 & 8.92480 & 35.737667 & 15 &  2.625349  \\
$6$ &45.372677& 1.254055 & 0.000002 & 0.000002 & 1.096524 & 21.724331 &  19.48275 & 5.11502 & 7.98101 & 6.26888 & 36.656037 & 9 & 2.230422 \\
$7$ &0.000002 & 1.401838 & 0.000001 & 0.000006 & 0.153699 & 20.237288 & 13.61419 & 10.26772 & 5.54050 & 1.55635 & 35.332118 & 5 &1.734501\\
$8$ &0.000546 & 1.132153 & 1.099589 & 0.000001 & 0.007597 & 0.000015 & 13.44260 & 4.68079 & 7.65450 & 4.68389 & 7.364656 & 2 & 1.392492 \\
$9$ &0.000005 & 0.425008 & 0.000018 & 0.000000 & 99.025827& 133.025276 & 5.02466 & 17.57855 & 6.32712 & 10.48183 & 11.919507 & 25 &  3.766443\\
\hline
\multicolumn{7}{r|}{Averages}&10.2906 & 9.52073 & 5.13994 & 4.93273 & 42.60624 & 17 & 2.907602\\
\hline
\end{tabular}
\label{TR}
\end{sidewaystable}

\newpage

\end{document}